\documentclass[12pt]{article}
\usepackage{color}
\usepackage{latexsym}
\usepackage{amsmath}
\usepackage{amsfonts}
\usepackage{amssymb}
\usepackage{indentfirst} 
\numberwithin{equation}{section}
\newcommand{\be}{\begin{equation}}
\newcommand{\ee}{\end{equation}}
 
\newcommand{\mR}{{\mathbb R}}

\newcommand{\mK}{{\mathcal K}}

\newcommand{\sign}{\textrm{sign}}

\newcommand{\tr}{\textrm{tr}}
\newcommand{\mA}{{\mathcal A}}

\newcommand{\bx}{{\bf{x}}}

\newcommand\braket[2]{\langle{#1}|{#2}\rangle}
\title{Global propagation  of massive quantum fields \\   in  the  plane gravitational  waves\\  and  electromagnetic  backgrounds }
\author{
\\ 
\vspace*{0.3cm}
K. Andrzejewski\footnote{e-mail: krzysztof.andrzejewski@uni.lodz.pl (the corresponding author)} \quad and  \quad
P. Kosi\'nski\footnote{e-mail: piotr.kosinski@uni.lodz.pl}\\
\vspace*{0.3cm}
\\
\small  Faculty of Physics and Applied Informatics,  \\ \small
University of  Lodz\\ \small Pomorska 149/153,
90-236, Lodz, Poland\\
}
\date{}
\begin{document}
\maketitle 
\begin{abstract} 
The   behavior of  massive quantum fields  in the   general  plane  wave spacetime   and  external,   non-plane, electromagnetic waves   is studied. The  asymptotic conditions, the ``in" (``out")  states    and  the  cross sections are  analysed. It is  observed  that,    despite of the  singularities encountered,  the    global form  of  these states    can be obtained:  at the singular points  the Dirac delta-like  behavior emerges and there is a discrete change of  phase of the wave function  after passing  through   each  singular point.    
 The relations between these phase corrections and local charts are discussed.   Some  examples  of waves of infinite range (including   the circularly polarized ones) are presented for which the explicit form of solutions can be obtained.  All these results concern both the scalar  as well  as  spin one-half   fields;  in latter case   the change of  the spin polarization  after   the general sandwich wave has passed   is studied.
\end{abstract} 
\newpage 
\section {Introduction}
The interaction of quantum    fields with  the  gravitational backgrounds has been  studied 
intensively   for  various spacetimes  for  years   (let us only  mention a few monographs \cite{b1a}-\cite
{b1d}).   Recently,  such investigations concerning the  gravitational waves  became  particularly 
interesting due to  direct observations of the latter from    a pair of merging black holes \cite{b2}    or 
colliding neutron stars \cite{b3};  for the latter case  the    electromagnetic   waves (backgrounds)  as well 
as spin  aspects    seem also to be important.   Although  the analysis of the  quantum  phenomena  for  the general  spacetime  (in the presence of electromagnetic fields) 
is very difficult there are some  relatively simple configurations  of fields   which can capture  some  
relevant    features of the  interaction. For the gravitational waves such models are  provided by the  plane 
gravitational waves, which  are   exact solutions to  the Einstein equations.  Of course,   such  models  
possess  physical limitations;  however, they can be useful in the analysis of strong or short time 
processes        which can occur in many  astrophysical and cosmological  phenomena   providing  their, 
at least,    local or   qualitative   descriptions.  
\par 
The analysis of  propagation of  quantum fields  in  the plane gravitational  wave   goes back to the pioneering  papers \cite{b4a,b4b}, where   some  basic properties  of   the scalar     fields  were established for the plane  sandwich  waves   (e.g.  no  particles creation and vacuum polarization).  Then,  these investigations  were  extended  in Ref.   \cite{b5} to    scattering processes in  the  fine-tuned (i.e. with   special compact supports)  plane sandwich  waves.     Similar results for  massive spin one-half    fields,  but  only for   linearly polarized  sandwiches, were  obtained in \cite{b6} (see also \cite{b7}).
Although the plane  gravitational waves   are relatively  simple they exhibit some     surprising  properties, e.g.  non-existence of a spacelike  Cauchy hypersurface;  when  colliding they can produce   curvature singularities.   Moreover, in spite of  the fact that the classical motion in these  spacetimes  is completely  regular the basic  wave functions  are  singular.    This fact  was related, for  compactly supported waves \cite{b4a,b5}, to the focusing  behavior   of the classical trajectories.
\par  On the other hand, it turns out that a similar situation  occurs for some electromagnetic fields in the Minkowski spacetime which   belong to  the  class of the so-called crossed fields  (in  the terminology of Ref. \cite{b13}).  Since these fields    include non-plane electromagnetic waves  they  are interesting in the context of  singular optics and  ultra intense    laser  beams (where  the  plane wave  approximation or  even paraxial one   is not sufficient). The similarity between  both the fields is even more evident if one  takes into account the double copy  conjecture, a part of the colour-kinematic duality which on the quantum  level  concerns the problem of how scattering amplitudes in gravity can be obtained from those in the gauge theory (see among others  \cite{b8a}-\cite{b8f}).    In view  of this and   the above mentioned  astrophysical phenomena we find  interesting to  investigate  the behavior  of the basic wave functions   around the singular  points (in consequence, their   global form)  simultaneously for  the   gravitational and electromagnet fields (see   Sec. \ref{s4}); of course, both the fields  can be considered separately, i.e.  gravitational  or only electromagnetic  fields  in the Minkowski spacetime. To make our considerations as  general as  possible,   first,  in Sec.  \ref{s0} we discuss  asymptotic conditions   which allow  to define properly the ``in" and ``out" states  (and, consequently,  to  analyse    the scattering processes, see  Sec. \ref{s3})  for arbitrary  (not only sandwich)  profiles. Moreover, since  our aim is to find  the  global form of states we work in the  Brinkmann  (B) coordinates; however, to make  contact  with various results in the literature  and the weak field approach in the last part of Sec. \ref{s0} we   refer to the  Baldwin-Jeffery-Rosen (BJR) coordinates which are local  because of the above mentioned  singular points.
\par 
Since  the interaction of  the    scalar  fields with plane wave spacetimes    was explicitly discussed   mainly for the standard (defined  by  characteristic  functions) sandwiches, in  Sec. \ref{s7}  we give some  examples  of  wave pulses of infinite range  for which the wave functions and the cross sections  can be explicitly written down;     these examples  include circularly   polarized waves  (which seem physically more interesting  \cite{b2,b3,b9,b9b}) and  in a  suitable limit they  reduce to impulsive  waves  \cite{b10}.    Finally, in Sec. \ref {s8}   we consider  all the  above issues for the spin one-half    fields; moreover,  we discuss    the change of the spin polarization  after the  passage of the general  sandwich wave  (for  standard  linearly polarized sandwich waves  this fact was discussed, by means of the  Newman-Penrose  framework,   in  Ref. \cite{b7}). Conclusions and  possible further directions of investigations  are presented in Sec. \ref{s9}.      
\section{Asymptotic conditions and scattering  }
\label{s0}
\subsection{General discussion}
\label{s1}
The dynamic   of the  massive scalar field $\Phi$  in  the  gravitational $g_{\mu\nu}$ and electromagnetic $A_{\mu}$  backgrounds  is described by the Klein-Gordon (K-G)  equation 
\be
\label{e1}
D_\mu D^\mu\Phi-m^2\Phi=0,
\ee
where $D_\mu=\nabla_\mu-ieA_\mu$ and $ \nabla$ is the Levi-Civita connection of the metric $g$ (we skip here the term with the scalar curvature since it  is  zero for the   metrics considered below).   
Equivalently,  eq. \eqref{e1} can be rewritten in  the form
\be
\label{e2}
\frac{1}{\sqrt{|\det(g)|}}\tilde D_\alpha(\sqrt{|\det(g)|}g^{\alpha \beta}\tilde D_\beta\Phi)-m^2\Phi=0,
\ee
with  $\tilde D_\alpha=\partial_\alpha-ie A_\alpha $. 
\par
Let us  consider   two types of fields: the gravitational one defined by a pp-wave  metric\footnote{The signature is $(-,+,+,+)$ and  the bold  symbols refer to the
two-dimensional vectors.}    describing, in general,     null fluid solutions (in particular, the plane  gravitational waves)
\be
\label{e3}
g= \mK(u,\bx )du^2+2dudv+d \bx^T d\bx ,
\ee
where  $\bx$ refers to the $x^1,x^2$ coordinates, while  the function $\mK=\mK(u,\bx)$  does not depend on the $v$ coordinate;  the   electromagnetic background is defined by the following potential $\mA_\mu$ 
\be
\label{e4}
\mA_u=\mA(u,\bx),\quad  \mA_v=0,\quad  \mA_i =0,\quad  i=1,2;
\ee
which also  does not depend on $v$.   
It is well known that for the metric   $g$ defined by \eqref{e3}  the scalar curvature  vanishes and   it    describes a  vacuum solution to the Einstein equations iff   $\triangle_\perp \mK\equiv \partial_i\partial^i\mK=0$; similarly,  one can check that the potential  $\mA_\mu$    describes a  vacuum solution  of the Maxwell equations (also in the curved spacetime defined by $g$)  when $\triangle_\perp \mA=0$ (note  that this condition  does not depend on the form of $\mK$, in particular putting $\mK=0$ one  obtains an electromagnetic wave in the Minkowski spacetime).
\par
For such fields   the coefficients in  eq. \eqref{e1} do not depend on the $v$ coordinate, thus  we look for    solutions in the following  form 
 \be
 \label{e5}
 \Phi(u,v,\bx)=e^{ivk_v-\frac{im^2u}{2k_v}}\bar  \Phi(u,\bx) ,
 \ee
 where  $k_v$ is  a non-zero  constant related to the  conservation law of the momentum in these backgrounds.  
Substituting   \eqref{e5}  into eq.  \eqref{e2}  one arrives at the  ``time"-dependent  ($u$ plays the   role of time)  Schr\"odinger   equation for the function  $\bar  \Phi$  
\be
\label{e6}
i\partial_ u\bar  \Phi=-\frac{1}{2k_v} {\triangle_\perp }\bar  \Phi-\frac{k_v}{2} (\mK +\frac{2e}{k_v}\mA)\bar  \Phi.
\ee
 In order to analyse  quantum processes   we need the  solutions of eq. \eqref{e6}  as well as  an inner product between  the full states $\Phi$'s;  this  is,  in general, not  an easy task. 
\par   
 The situation slightly  simplifies under the  assumption that both $\mA$ and $ \mK$ are quadratic functions of $\bx$, i.e.   
\be
\label{e7}
\mK(u,\bx)=\bx^TK(u)\bx,
\ee
and 
\be
\label{e8}
\mA(u,\bx)=\bx^TA(u)\bx,
\ee
where $A(u)$ and $H(u)$ are continuous\footnote{In many places this assumption can be relaxed to include discontinuous  sandwich functions.} matrix functions (without losing  generality  assumed to be  symmetric ones). Recall  also here  that  the coordinates for which the metric $g$  is of the form  \eqref{e3}  with  \eqref{e7} are called the Brinkmann  (B) ones (see, e.g.,  \cite{b5,b11} and references therein  as well as  Sec. \ref{s6}).  
\par 
For such a choice of $\mA$ and $ \mK$ eq. \eqref{e6}  reduces  to, in general anisotropic and time dependent, quantum    linear oscillator which  has been  intensively studied  since the   classical papers \cite{b12a,b12b}.   
However,  as we  mentioned  above,    eq.  \eqref{e6}  is an auxiliary one  and  in order to analyse the processes in such backgrounds  one needs  an inner  product between  $\Phi$'s.  This is a  subtle point since the spacetime $g$  does  not possess a Cauchy surface (except  the  Minkowski case, $K=0$).  However, it turns out \cite{b4a,b5} that the   null surface $u=\textrm{const}$ can be a good substitute for it;  then the  inner  product  can be defined as follows
\be
\label{e9}
\braket{\Phi_1}{ \Phi_2}=-i\int_{\mR^3}(\Phi_1\partial_v \Phi_2^*- \Phi_2^*\partial_v\Phi_1) dvd\bx. 
\ee
Since  $A_v=0$ the above inner product does not contain electromagnetic potential. Moreover,  even for the Minkowski spacetime the inner product \eqref{e9}  is related to a   null hypersurface;  despite  this fact it is  physically interesting  due to its usefulness in the  intensive   studies  of the  electromagnetic beams (see \cite{b13} and references therein). Finally,  let us note that   the choice  of the   gravitational and  electromagnetic potentials \eqref{e7} and \eqref{e8}  includes many  interesting cases, e.g. the   exact  gravitational waves,   $\tr(K)=0$,  and non-plane electromagnetic waves, $\tr(A)=0$  (although  we do not exclude the null fluid solutions\footnote{Sometimes  the metric $g$ with  an arbitrary $K$ is  called the generalized plane gravitational wave. }  in our considerations).  
\par
From the above we see that in order  to analyse the K-G  equation   \eqref{e1},  with  the fields defined by \eqref{e7} and \eqref{e8},  one should find solutions to eq.  \eqref{e6}. To this end, let us consider  the  matrix  function $P$ which is a solution  of  the following system of the linear differential equations  
\be
\label{e10}
\overset{..}{P}=(K+\frac{2e}{k_v}A)P,
\ee
where dot refers to the derivative w.r.t. to $u$ (let us stress   that in the presence of the electromagnetic potential  $P$  depends on $k_v$; however, to simplify our notation we skip this subscript). A few remarks are in order. 
First,  one can impose  a condition  on the solutions of \eqref{e10}, namely
\be
\label{e11}
\dot P^TP=P^T\dot P .
\ee
 Next,   following  the reasoning presented in  Ref.  \cite{b5}  we conclude  that  when   $\tr(K+\frac{2e}{k_v}A)\leq 0$ and  
 $K+\frac{2e}{k_v}A\neq 0$ (i.e. except the trivial case)  there exists at least one point (the so-called singular point) where the determinant of the  matrix  $P$ vanishes (note that when    the electromagnetic field is turned  on then this singular  point depends on $k_v$).  Such a situation  appears in the  most interesting case  when   $g$ is an exact   gravitational wave  and the  electromagnetic field  is a non-plane electromagnetic wave; another example  of such a situation is provided  by   $A=0$ and   $g$ satisfying  the weak energy condition.  
\par Let us now  define   the antiderivative  matrix  $S$ as  $\dot S=(P^TP)^{-1}$; then the field
\be
\label{e12}
\bar   \Phi^{k_v,{\bf k}}(u,\bx)  =\frac{1}{\sqrt{2|k_v|(2\pi)^3|\det(P(u))|}}e^{i{\bf k}^TP^{-1}(u)\bx+i\frac{k_v}{2}\bx^T\dot P(u)P^{-1}(u)\bx-\frac{i}{2k_v}{\bf k}^TS(u){\bf k}},
\ee
depending on the   constants $k_v,{\bf k}$,   satisfies eq. \eqref{e6}   and leads, by virtue of eq.  \eqref{e5},  to the  solution  $\Phi^{k_v,{\bf k}}$ of  the K-G equation  \eqref{e1};  at least for all $u$ where   $P$ is  invertible  (say, before the singular point). At the singular point we have   three not  well defined terms in \eqref{e12}:  $\det(P)$ in the denominator, $P^{-1}$ and  $S$ in the exponential factor.  However, we will show below that  all together  combine to  a correct, in the distributional sense, behavior of the state \eqref{e12}  as $u$ tends to  the singular point; namely, one obtains the one or two-dimensional  Dirac delta function (depending on the rank of  $P$ at the singular point).  In view of this and the fact that  the field defined by the right hand side of   \eqref{e12} is a solution to the  K-G equation  in all nonsingular intervals, one  can try to treat    the field  $\Phi^{k_v,{\bf k}}$   globally  (i.e. for all real $u$). Then, however,   a problem  arises since  the antiderivative  $S$ is not uniquely determined  (there can appear   a constant matrix  because of the disconnected domain); thus  after passing  through  the singular point  an  arbitrary   phase factor   can  appear. In consequence, the behavior of $\Phi^{k_v,{\bf k}}$  around the singular points calls for a careful analysis.  In what follows we  show that  beyond  the singular point the dynamic of   the states  is uniquely determined, though there appears a discrete phase correction (depending  on the rank of  $P$ at the singular point only).    Such a  jump of   the phase is similar to the one  appearing  for  the propagator (Green function)   passing  through a caustic point; for example, it  has been     extensively  studied  for the  harmonic oscillator \cite{b14a,b14b},  as well as  encountered  in various  optical, molecular, nuclear   contexts (see  \cite{b15} and  references therein).  Such an effect   has been  also  noted   for  the Green functions \cite{b16,b18a,b17}   of  the plane gravitational waves  for which geodesics exhibit  conjugated points (caustics);  in consequence,  some very  interesting   results concerning  the  vacuum polarization  and refractive index  were obtained   \cite{b16}-\cite{b18c}. It is worth to notice that  for the states under considerations the initial conditions,  see eqs.    \eqref{e16} below,  are   different  than the ones encountered  for   the  propagator  (Green functions);  cf. eqs. \eqref{e37aa}  and   the further discussion in  Sec. \ref{s5}.   Moreover, the conjugated points are not  generic for the vacuum solutions which are our  main interest. 
\par
Concluding these  preliminaries let us note that the  fields $\Phi^{k_v,{\bf k}}$  where $k_v\in \mR\backslash\{0\}, {\bf k}\in\mR^2$   form, for a fixed $u$, a  complete and  orthonormal  set   of functions   with respect to the inner  product \eqref{e9}       (including  the electromagnetic case   when   $P$ depends on $k_v$).  In our convention,  the states with  $k_v>0$  have  negative norms  and correspond to antiparticle states  (cf.   \eqref{e17}); moreover,  they are orthogonal to the ones with $k_v<0$.
 When we consider the  gravitational backgrounds only  (without the electromagnetic  fields)  then $P$ does not depend on $k_v$ and one can replace  $\Phi^{k_v,{\bf k }}$  with  $k_v>0$, by   $ {\Phi^{k_v,{\bf k}}}^*$ with $k_v<0$. 
 \subsection{Asymptotic conditions}
 \label{s2}
\par  
In order to analyse  the  interaction of    scalar  fields  with  the gravitational as well as electromagnetic backgrounds (defined by \eqref{e7} and \eqref{e8})  we   start with the  states  which  form  at minus (plus)  null infinity the plane wave solutions of the  K-G equation. Since we  do not restrict ourselves  to the  backgrounds  which are equal  to zero for sufficiently  large $|u|$, i.e.    with  compact supports  (cf. gravitational  sandwiches \cite{b5})   we start with a  discussion of  the asymptotic conditions which should be  imposed on the   matrices  $A$ and $K$  to make  such a situation possible.  Applying    Lemma 1.2 from  Ref.  \cite{b19} (which generalizes Theorem 2 in \cite{b20})     we conclude that for the functions\footnote{We do not specify the  matrix norm,  one can take e.g.  the Frobenius norm.}     $u||K(u)||$ and $u||A(u)||$ belonging to   $L^1(\mR)$     there exist  the solutions  $P_{in}$  and $P_{out}$ of eq. \eqref{e10}   satisfying     the following initial conditions 
 \be
 \label{e16}
\dot P_{in}(-\infty)=0,\quad  P_{in}(-\infty)=I, \quad  \dot P_{out}(\infty)=0, \quad P_{out}(\infty)=I;
 \ee
 note that these conditions are compatible with  \eqref{e11}; moreover, if $K$ and $A$  are even functions    we can put $P_{out}(u)=P_{in}(-u)$. 
 In spite of   asymptotic vanishing of the profiles $K$ and $A$,    the general solution of eqs. \eqref{e10}  does not, in general,     behave  near  plus (minus) infinity as one would expect.  In   fact, even  on the classical level   the straight-line geodesics behavior  is provided by  the following asymptotic conditions at infinities   
\be  
\label{e14}
P(u)\simeq   P_1u+P_2,
\ee
 where $P_1,P_2$ are constant matrices  and  $\simeq$   denote the  relation that holds asymptotically (or in a sufficiently small neighborhood).
To enforce the conditions \eqref{e14}  slightly stronger  assumptions   concerning the  profiles  are needed;   namely, the functions   $u^2||K(u)||$ and $u^2||A(u)||$   should belong to  the   $L^1(\mR)$ space\footnote{For the  diagonal profiles this fact follows from Theorem 3 \cite{b20} (see also the third footnote therein); however,   the proof of this theorem can be  directly extended to the arbitrary case -- in the same way  as    Theorem 2 in \cite{b20} is extended by  Lemma 1.2  in   \cite{b19}.}. 
In view of this  we have two possibilities:  either   
\be
\label{e15} 
\lim_{u\rightarrow \pm\infty }  u^2K(u)=0, \quad \lim_{u\rightarrow \pm\infty }  u^2A(u)=0;
\ee
or  the limits  on the left-hand side of \eqref{e15} do not exist.  In what follows we will focus  on the  first case, since the integrable functions for which the above  limits do not exist  have rather  peculiar  behavior at  infinities, especially if we take into account  that they should be related to  the gravitational and  electromagnetic fields. Moreover, they can lead to the states for which the asymptotic form does  not coincide with the one   for the Minkowski spacetime\footnote{Namely, instead of eq. \eqref{e20a} we have  $W=\dot P_{in}(\infty)-\lim_{u\rightarrow \infty}\dot P_{out}^T(u)P_{in} (u)$. Although the right-hand side  is well defined, it is difficult  to check whether $W$ is invertible; moreover,  even if $W$ were invertible  then the  asymptotic form of the  state would  be different from \eqref{e17}.}.
These peculiarities can be seen also at the classical  level.  Namely, it turns out that   the geodesic equation (or the  Lorentz equation)  for  the longitudinal  coordinate $v$ in   the discussed backgrounds   can be once integrated yielding $\dot v=-\frac 12 \dot \bx^2-\frac 12 \bx^TK\bx+c_0$. On the other hand,    $\dot \bx  $  tends to a constant while $\bx$ is (in general)  proportional to $u$ when $u$ tends to infinities. In consequence,   $\lim_{u\rightarrow \infty}\dot v(u)\simeq -\frac 12 \lim_{u\rightarrow\infty}u^2K(u)+const$ and  the longitudinal  velocity may be not well defined. 
 \par 
As we  have noted above  in many physically interesting  cases there   appear  singular  points  where the determinants  of $P_{in}$  and $P_{out}$  vanish and  consequently the behavior of the states should be carefully analysed.  Let us  first assume that there is only one point $u_{in}$ such that $\det(P_{in}(u_{in}))=0$ and one  point $u_{out}<u_{in}$ such that   $\det(P_{out}(u_{out}))=0$ (some  criteria enforcing such a situation can be found in \cite{b21}); the case of more  singular points will be discussed latter (cf. Sec. \ref{s5}).     In such a case   for $u\in (u_{out},u_{in})$ there are two orthonormal sets of the  states determined by   $\Phi_{in}^{k_v,{\bf k}}$ and $\Phi_{out}^{l_v,{\bf l}}$.
 \par  Before we proceed let us   make  some useful  observations.  To this end let us  define\footnote{$W$ can be defined for arbitrary two solutions of eq. \eqref{e10}.  } 
\be
\label{e18}
W\equiv W(P_{in},P_{out})= P_{out}^T\dot P_{in}-\dot P_{out}^TP_{in}  .
\ee
Then  $\dot W=0$,  which  implies that $W$ is a constant  matrix and, by virtue  of \eqref{e11}, the following relation hold for $u> u_{out}$
\be
\label{e19}
(S_{out}(u)-S_{out}(u_0))W=P_{out}^{-1}(u) P_{in}(u)-P_{out}^{-1}(u_0) P_{in}(u_0),
\ee
as well as for  $u< u_{in}$
\be
\label{e20}
(S_{in}(u)-S_{in}(u_0))W^T=-P_{in}^{-1}(u) P_{out}(u)+P_{in}^{-1}(u_0) P_{out}(u_0),
\ee 
where $u_0$  is such that $u_{in}<u_0<u_{out}$. 
Next,  let us note that 
\be
\label{e20a} 
W=\dot P_{in}(\infty), \quad W^T=-\dot P_{out}(-\infty).
\ee
Indeed, by   virtue of  \eqref{e16}, \eqref{e14}, \eqref{e15} and L'Hospital's rule  we have 
\be
\label{e21}
\lim_{u\rightarrow \infty}\dot P^T_{out}(u)P_{in}(u)=\lim_{u\rightarrow \infty}u\dot P^T_{out}(u)P_1=-\lim_{u\rightarrow \infty}u^2(K(u)+\frac{2e}{k_v}A(u))P_1=0,
\ee
and similarly for the second case in eqs.  \eqref{e20a}.
In what follows  we assume that $W$ is an invertible matrix; this   assumption has a physical interpretation that the final momenta are not linearly dependent and  the cross section is well defined, see below  (in particular, the final momenta cannot vanish after scattering).   
 \par
Now, we  are in the position to analyse the   asymptotic  behavior of $\Phi_{in}^{k_v,{\bf k}}$   defined by $P_{in}$    near minus infinity.  To this end   let us note that  by virtue of   \eqref{e14}, \eqref{e20}  and  \eqref{e20a}     the antiderivative  behaves at minus infinity  as  $S_{in}(u) \simeq  uI+C_{in}$  where  $C_{in}$ is  a constant matrix.  However, the   matrix     $C_{in}$ can be eliminated by a suitable choice of $S$ (since the constant in the   antiderivative had not   been  used yet). In consequence,  with such a choice of $S$   one has the following asymptotic behavior of the ``in" states  near minus infinity
\be
\label{e17}
\Phi_{in}^{k_v, {\bf k}}(u,v,\bx)\sim  e^{-\frac{im^2u}{2k_v}+ik_vv+i{\bf k}^T\bx-\frac{i}{2k_v}{\bf k}^T{\bf k}u}=e^{i(\vec{k}^T\vec{x}-k_0t)};
\ee
in our convention,  for the Minkowski spacetime, we have the   identifications  $u=\frac{1}{\sqrt 2}(x^3-x^0)$  and $k_{v}=\frac{1}{\sqrt 2}(k_3-k_0)$; moreover,  $k_v<0$ iff $k_0>0$.  Let us stress that in order to obtain \eqref{e17} not only \eqref{e16} but also the condition \eqref{e15} was  used (this  coincides with the classical   straight-line geodesics analysis). The same analysis can be performed  for the states $\Phi_{out}^{k_v, {\bf k}}$,  defined by  $P_{out}$, at plus infinity. 
\par  Moreover, for  further purposes  let us analyse the  behavior of the matrix function $P_{in}(u)$ near the singular point  (the  reasoning  presented in this paragraph is similar to the one in Ref. \cite{b17}; however,  we do not restrict ourselves to the caustic points and the initial  conditions are different).   First, let us consider the degenerate case when  $P_{in}(u_{in})=0$ i.e. the rank is zero. Then, in  a small neighborhood of $u_{in}$ we have  $P_{in}(u)\simeq (u-u_{in})\dot P_{in}(u_{in})$. In consequence, to   the lowest order   $\det( P_{in}(u))\simeq (u-u_{in})^2\det\dot P_{in}(u_{in}) $.  On the other hand 
\be
\label{e17b}
W= P_{out}^T(u_{in})\dot P_{in}(u_{in}),
\ee
 thus due to invertibility of  $W$ we find  that   $\det \dot P_{in}(u_{in})\neq 0$ and  $\det  P_{in}(u)$  does  not change the sign  after  passing  through  the point $u_{in}$.
 The situation is quite different if the rank of  the matrix $P_{in}(u_{in})$ is one  (for example, this holds  for the vacuum linearly polarized waves  with  non-vanishing profiles; in fact, one component of $P_{in}$  is concave while  the second  one is   convex    before $u_{in}$, see eq. \eqref{e10},    thus only one component  of $P_{in}$ can vanish at $u_{in}$).  Expanding  the function $\det (P_{in}(u))$ we arrive,  after some computations,   at the first order equality 
\be
\label{e22}
\det (P_{in}(u)) \simeq (u-u_{in}) \tr (\dot{ P}^{-1}_{in}(u_{in})P_{in}(u_{in}))\det(\dot P_{in}(u_{in})).
\ee
Thus making non-restrictive   assumption\footnote{The singular points of $P_{in}$ and  $\dot P_{in}$  are isolated.}  that $\det(\dot P_{in}(u_{in}))\neq 0$  (which  holds, for example,  in the above mentioned case of linearly polarized waves) we find  that the rank of the matrix under the trace is also one. Moreover,  by virtue of \eqref{e11} it is  a symmetric matrix; in consequence, it has the non-vanishing trace. Concluding, in this case   $\det (P_{in}(u)) $ changes the sign  after  passing  through    the point $u_{in}$. The above  results will be useful  in what follows.  
\subsection{The quantum scattering }
\label{s3}
Let us  analyse the scattering   of the  scalar   particle states  in the  discussed backgrounds (in particular,  the quantum cross section).  To this end we compute the transition amplitude between ``in" and ``out" states. We will  use the  $u=u_0$ ($u_{in}<u_0<u_{out}$)  null hypersurface to compute the inner  product (let us stress that, in contrast to the  fine-tuned  sandwiches, for the    general continuous pulses  the point $u_0$ cannot be chosen   in the region where $P_{out}$ is the   identity matrix, cf. Ref.  \cite{b5}). 
First, we note that $\braket{\Phi_{in}^{k_v,{\bf k}}}{\Phi_{out}^{l_v,{\bf l}}} =0$ for $k_v<0$ and $l_v>0$, i.e   ``in" vacuum and ``out" vacuum can be identified, since   the positive-energy ``in"  modes do not develop   negative-energy  ``out" mode parts;  there is no particles creation  (this holds for the massive case $m\neq 0$; in the massless case the situation  may be different, see  \cite{b22a,b22b} and references therein).
Next, for $k_v,l_v<0$ we find 
\be
\label{e23}
\braket{\Phi_{in}^{k_v,{\bf k}}}{\Phi_{out}^{l_v,{\bf l}}}(u_0)=\frac{\delta(k_v-l_v)}{(2\pi)^2}\frac{\sqrt{|\det(P_{in}(u_0))|}}{{\sqrt{|\det(P_{out}(u_0))|}}}e^{\frac{i}{2l_v}({\bf l}^TS_{out}(u_0){\bf l}-{\bf k}^TS_{in}(u_0){\bf k})}\int e^{i{\bf y}^T M(u_0){\bf y}+i{\bf z}^T{\bf y}} d{\bf y},
\ee
where ${\bf z}={\bf k}-P_{in}^T(u_0)(P_{out}^{T}(u_0))^{-1}{\bf l}$ and the matrix  $M(u)$  defined  for $u>u_{out}$  by
\be
\label{e24}
M(u)=\frac{k_v}{2}(P_{out}^{-1}(u)P_{in}(u))^T W,
\ee
is a  symmetric  one, $M^T=M$. 
Using  the well known  formula valid  for a nonsingular\footnote{For the singular matrix $M$ the right hand side of  eq.  \eqref{e25} reduces to the one or two dimensional  (depending on the rank of $M$)  Dirac delta function  multiplied  by  suitable coefficients.}     and  symmetric two-dimensional matrix
\be
\label{e25}
\int  e^{i{\bf y}^T M{\bf y}+i{\bf z}^T{\bf y}} d{\bf y}=\frac{\pi}{\sqrt{|\det(M)|}}e^{\frac{i\pi}{4}\\\sign(M)}e^{-\frac{i}{4}{\bf z}^TM^{-1}{\bf z}},
\ee
where $\sign(M)$ is the signature of $M$ (the number of positive eigenvalues minus the number of negative eigenvalues) one obtains 
\be
\label{e26}
\braket{\Phi_{in}^{k_v,{\bf k}}}{\Phi_{out}^{l_v,{\bf l}}}(u_0)=\frac{\delta(k_v-l_v)}{2\pi|k_v|\sqrt{|\det(W)|}}e^{\frac{i\pi}{4}\sign(M(u_0))}e^{-\frac {i}{4} {\bf z}^TM^{-1}(u_0){\bf z}}e^{\frac{i}{2l_v}({\bf l}^TS_{out}(u_0){\bf l}-{\bf k}^TS_{in}(u_0){\bf k})}.
\ee
Eq. \eqref{e26} simplifies   when $K$ and $A$ are even matrix functions and $u_0=0$.  Then  
\be
\label{e27}
-{\frac i 4 {\bf z}^TM(0)^{-1}{\bf z}}=\frac{-i}{2k_v}({\bf {\bf k}-{\bf l})}^TW^{-1}({\bf k}-{\bf l}),
\ee
and $W=2P_{in}^T(0)\dot P_{in}(0)$;  moreover,  $S_{out}(0)=-S_{in}(0)$.
\par 
It remains to find the signature of $M(u_0)$. Since the behavior of $M$ near plus infinity  is of  the form  $M(u)\simeq \frac{k_v}{2}uW^TW$   one obtains  (for the particle state, i.e.  $k_v<0$)   $\sign(M(u))=-2$ for $u>u_{in}$.  Let us consider two cases.    First, the   degenerate case,  i.e.   $P_{in}(u_{in}) =0$; then the rank   $M(u_{in})$  is zero and    expanding it in a neighborhood of $u_{in}$, we have $M(u)\simeq \frac{k_v}{2}(u-u_{in})\dot P_{in}^T(u_{in})\dot P_{in}(u_{in})$. On the other hand  taking into account that  $W$ is invertible we get    $\det(\dot P_{in})(u_{in})\neq 0$; thus both the  eigenvalues of $M(u)$ change the sign after passing through    $u_{in}$. In consequence,  $\textrm{sign}(M(u_0))=2$ and an additional factor in \eqref{e26}  emerges.  Next, let us consider the case when the rank of  $P_{in}(u_{in})$ is one. Then,  by virtue of our previous considerations, we  can assume that the determinant of  $P_{in}(u)$, and consequently of $M(u)$, changes the sign at $u_{in}$.  In consequence,   only    one eigenvalue  function  of $M(u)$ becomes positive  for $u<u_{in}$  implying  that  $\sign(M(u_0))=0$. For  further use   let us also  note that  for $u>u_{in}$  the difference    $\sign(M(u_0))-\sign(M(u))$ equals $2$ (or $4$) if    the rank of $P_{in}(u_{in})$ is one (or zero). 
\par 
In view of the above  we  can easily  find  the transition amplitude from an ``in" one-particle state with $(k_v,{\bf k})$ to  an ``out" one-particle state $(l_v,{\bf l})$. In fact, since there is no particle creation   both vacua can be identified, and the following relation holds
\be
\label{e28}
\braket{\textrm{out},l_v,{\bf l}}{k_v,{\bf k},\textrm{in}}=\braket{\Phi_{in}^{k_v,{\bf k}}}{\Phi_{out}^{l_v,{\bf l}}}(u_0),
\ee
thus
\be
\label{e29}
|\braket{\textrm{out},l_v,{\bf l}}{k_v,{\bf k},\textrm{in}}|=\frac{\delta(k_v-l_v)}{2\pi |k_v|\sqrt{|\det(W)}|} .
\ee
Following the standard reasoning  (and  taking into account the normalization related to infinite range of the $v$ coordinate,   see  Ref. \cite{b5})  we obtain  the quantum cross section 
\be
\label{e30}
d\sigma= \frac{1}{l_v^2{|\det(W)}|} d{\bf l}.
\ee
Finally,  let us note that,  by virtue  of  eqs.  \eqref{e20a},  it coincides with the classical differential cross section obtained in    \cite{b23} (in full agreement with    the discussion presented   in Ref. \cite{b5}).
 \section{Global states}
  \label{s4}
  \subsection{The first approach}
  Now, let us  analyse the behavior of  the ``in" states at   and beyond the singular point.   In principle, the behavior of $\Phi_{in}$'s at $u_{in}$ can be obtained by carefully  taking the  limit $\lim_{ u\rightarrow u_{in}^-}\Phi_{in}^{k_{v},{\bf k}}$.  However, since our aim is to find   the  global form  of  the states (i.e. also  beyond the singular point)  such a reasoning  will be considered later and then  compared to our  global result.  To find the  global form of the ``in"  states  we  follow two approaches.  One is based on the local  ``out" states  (assuming as above that there is one singular point and  $u_{out}<u_{in}$); another approach,  based on the evolution operator,  is discussed in the next subsection (it  works also in   the case  of more singular points).  
\par The main idea of the first approach  is that the   form  of the state $\Phi_{in}^{k_{v},{\bf k}}$ (defined for $u<u_{in}$)  at and beyond the singular point $u_{in}$     can be obtained  by means of the ``out" states which are well defined around  $u_{in}$ (since $u_{out}<u_{in}$),  and form also an orthonormal basis.   Namely,  in this approach the state  $\Phi_{in}^{k_v,{\bf k}}$  is  extended     to   $\tilde \Phi_{in}^{k_v,{\bf k}}$,  for all $u>u_{out}$,     by the    relation  
 \be
 \label{e31}
 \tilde \Phi_{in}^{k_v,{\bf k}}(u,v,\bx)=\int  \Phi_{out}^{l_v,{\bf l}}(u,v,\bx) \braket{\Phi_{in}^{k_v,{\bf k}}}{\Phi_{out}^{l_v,{\bf l}}}(u_0) dl_v d{\bf l}.
 \ee
 Let us analyse the above  solution of the  K-G equation for the particle state $k_v<0$.   Substituting \eqref{e5}, \eqref{e12} and  \eqref{e26}  into \eqref{e31}    we  are faced with  the  following integral   
 \be
\label{e32}
 \tilde \Phi_{in}^{k_v,{\bf k}}(u,v,\bx)=(...)\int 	e^{\frac{-i}{2k_v}{\bf z }^TW^{-1}(P_{in}^{-1}P_{out})^T({u_0}){\bf z}+i{\bf l}^TP_{out}^{-1}(u){\bf x}+\frac{i}{2k_v}{\bf l}^T(S_{out}(u_0)-S_{out}(u)){\bf l} }d{\bf l},
 \ee
 where $(...)$  denotes  regular terms skipped  in the  intermediate steps.  Using the identity \eqref{e19},   the definition of ${\bf z}$ and  making the  substitution  ${\bf l}=P^T_{out}(u){\bf q}$ under integral we arrive at  the expression 
  \begin{align}
  \label{e33}
 \tilde \Phi_{in}^{k_v,{\bf k}}(u,v,\bx)=&(...)|\det(P_{out}(u))|e^{\frac{-i}{2k_v}{\bf k}^TW^{-1}(P_{in}^{-1}P_{out})^T(u_0){\bf k}}\nonumber \\
&\cdot \int 	e^{i(\frac{1}{k_v}P_{out}(u)(W^T)^{-1}{\bf k}+\bx )^T{\bf q}+i{\bf q}^T  \tilde M(u){\bf q} }d{\bf q} \, ,
 \end{align}
 where the matrix $\tilde M(u) = -\frac{1}{2k_v} P_{in}(u)W^{-1}P_{out}^T(u) $ is a symmetric one.  
\par 
Now we  are ready to  analyse the state  $\tilde \Phi_{in}$. First, let us    assume that   $u>u_{out}$ and $u\neq u_{in}$. Then  $\tilde  M(u)$ is nonsingular and  $\tilde M=-\frac{1}{k_v^2}P_{in}M^{-1}P_{in}^T$, thus  $\sign(\tilde M)=-\sign(M)$. Using these facts and  the  formula \eqref{e25}  we find
\begin{align}
\label{e34}
\tilde \Phi_{in}^{k_v,{\bf k}}(u,v,\bx )&=\frac{e^{\frac{i\pi}{4}(\sign(M(u_0))-\sign(M(u)))}}{\sqrt{-2k_v(2\pi)^3}\sqrt{|\det(P_{in}(u))|}}{e^{-\frac{im^2u}{2k_v}+ik_vv+\frac{ik_v}{2}\bx^T\dot P_{in}(u)P_{in}^{-1}(u)\bx+i{\bf k}^TP_{in}^{-1}(u)\bx}}\nonumber \\
& e^{-\frac{i}{2k_v}{\bf k}^T(S_{in}(u_0)+W^{-1}(P_{out}^T(u_0)(P_{in}^{-1}(u_0))^T-P_{out}^T(u)(P_{in}^{-1}(u))^T){\bf k}}.
\end{align}
It is worth to notice that  the above formula  does not depend on the choice of $u_0$, as one can expect. Let us consider two cases.  
\par 
First,  assume that     $u\in (u_{out}, u_{in})$. Then     $\sign(M(u))=\sign(M(u_0))$;  moreover,  using \eqref{e20}  we can  eliminate $P_{out}$   in favor of $P_{in}$. As a result   one obtains   the expected  relation $\tilde \Phi_{in}(u,v,\bx)= \Phi_{in}(u,v,\bx)$,    i.e.  $\tilde \Phi_{in}$ is  indeed  an extension of the field  $\Phi_{in}$ defined for $u<u_{in}$. 
\par 
Next,  let us analyse the case $u>u_{in}$.  The last  exponential factor in  \eqref{e34}, corresponding to $\tilde S_{in}$,   is uniquely determined. Moreover, since  
\be
\label{e34a}
S_{in}(u)= W^{-1}(P_{out}^T(u_0)(P_{in}^{-1}(u_0))^T-P_{out}^T(u)(P_{in}^{-1}(u))^T)+S_{in}(u_0),
\ee 
 for $u<u_{in}$,  we   can identify  $\tilde S_{in}(u)=S_{in}(u)$ also   for $u>u_{in}$  -- despite of the disconnectedness of the  domain there  is no additional constant matrix in the antiderivative  of $(P_{in}^TP_{in})^{-1} $.   In view of this  eq.  \eqref{e34} can  be written as follows 
 \be
\label{e35}
\tilde \Phi_{in}^{k_v,{\bf k}}(u,v,\bx)=e^{\frac{i\pi}{4}(\sign(M(u_0))-\sign(M(u))} \Phi_{in}^{k_v,{\bf k}}(u,v,\bx) ,
\ee
  where  $\Phi_{in}^{k_v,{\bf k}}$   is considered for $u\neq u_{in}$ (with the antiderivative $S_{in}(u) $   given by the matrix function  occurring on  the right hand side of eq.  \eqref{e34a}).
In consequence,     there is   a discrete change of the phase beyond the singular point    related  only to  the change of  the signature of $M$. Namely,  by virtue of our previous considerations   (see the discussion  below \eqref{e27}) for  $ u>u_{in}$    the  $\sign(M(u_0))-\sign(M(u))$ equals $2$ (or $4$) if    the rank of $P_{in}(u_{in})$ is one (or zero). Thus, for the particle state $k_v<0$, beyond  the point $u_{in}$ the additional   factor  $i$  (or   $-1$, respectively)  appears.  
\par 
Finally, let us consider  the behavior of $\tilde \Phi_{in}^{k_v,{\bf k}}$ at $u_{in}$.   At this point  the rank of $P_{in}(u_{in})$ is  zero or one. Let us start with the case $P_{in}(u_{in})=0$.  Then the second term under the integral in \eqref{e33}  drops out and using  \eqref{e17b}   we arrive,  after some computations, at the Dirac delta function
\begin{align}
\label{e36}
\tilde \Phi_{in}^{k_v,{\bf k}}(u_{in},v,\bx)=&\frac{ ie^{\frac{-im^2u_{in}}{2k_v}+ik_vv }e^{\frac{i}{2k_v}{\bf k}^T\left (\dot P_{in}^{-1}(u_{in})\dot P_{out}(u_{in})-P_{in}^{-1}(u_0)P_{out}(u_0)-S_{in}(u_0)W^T\right)(W^{-1})^{T}{\bf k} }}{\sqrt{-4k_v^3\pi |\det (\dot P_{in}(u_{in}))|}} \nonumber \\
&\cdot \delta^{(2)}(\bx +\frac{1}{k_v}P_{out}(u_{in})(W^T)^{-1}{\bf k});
\end{align}
note that  \eqref{e36} does not depend on the choice of $u_0<u_{in}$. Now, let us us consider the case when   the rank of $P_{in}(u_{in})$ is one. Since the matrix $P_{in}W^{-1}P_{out}^T$ is symmetric  there exists an orthogonal  matrix $R$ such that 
\be 
\label{e36a}
R^T P_{in}(u_{in})W^{-1}P_{out}(u_{in})^TR=D,
\ee where  $D$ is a  diagonal matrix  with  the (say) second  element equal to  zero.   Then,  in the new coordinates   ${\bf y}=R^T{\bf x}$,  the  exponential factor under the  integral factorize: the factor  related to the $y^1$ coordinate can be computed by means of the one dimensional counterpart of the  formula \eqref{e25}, the  second one, related to $y^2$, yields  the one-dimensional  Dirac delta function $\delta(y^2+\frac{1}{k_v}(R^TP_{out}(u_{in})(W^T)^{-1}{\bf k})^{(2) })$. 
\par 
 Now, let us   compare    the form   of $\tilde \Phi ^{k_{v},{\bf k}_{in}}$ at $u_{in}$, obtained above,    with the following limit $\lim_{ u\rightarrow u_{in}^-}\Phi_{in}^{k_{v},{\bf k}}$. As we noted above  $\Phi_{in}$'s contain a few terms related to $P_{in}^{-1}$    (which become meaningless at $u_{in}$); moreover, we expect the  Dirac delta behavior (see \eqref{e36}); thus the distributional character  of that  limit  should be  taken into account. In view of this,  we first take the Fourier transform (w.r.t {\bx }) of the state,   then  perform the limit and, finally, take the inverse Fourier transform.  As previously,   we make this procedure for  $k_v<0$ and both cases of  the rank of $P_{in}(u_{in})$  separately.  Let us start with the rank zero  case. Then    for $u<u_{in}$   and  sufficiently close to  $u_{in}$    one has $\det( \dot P_{in}(u))\neq 0$  (since $\det(\dot  P_{in}(u_{in}))\neq 0$)  and thus  $\sign(\frac{k_v}{2}\dot P_{in}(u)P_{in}^{-1}(u))=\sign(\frac{k_v}{2}P_{in}(u)\dot P_{in}^{-1}(u))= \sign(\frac{k_v}{2}(u-u_{in})I)=2$. In consequence, for such $u$,   by   virtue of eqs.  \eqref{e5},\eqref{e12}  and   \eqref{e25},  we have   
\begin{align}
\label{e36b}
\int e^{-i\bf{q}^T\bx }\Phi_{in}^{k_{v},{\bf k}}(u,v,\bx) d{\bx}=&\frac{e^{\frac{-im^2u}{2k_v}+ik_vv +\frac{\pi i}{2}}}{\sqrt{-4k_v^3\pi |\det( \dot P_{in}(u))|}}e^{-\frac{i}{2k_v}{\bf q}^TP_{in}(u)\dot P_{in}^{-1}(u){\bf  q}+\frac{i}{k_v}{\bf k}^T\dot P_{in}^{-1}(u){\bf q}}\nonumber\\
&\cdot e^{\frac{-i}{2k_v}{\bf k}^T\left(S_{in}(u)+P_{in}^{-1}(u)(\dot P_{in}^{-1}(u))^T\right){\bf k}},
\end{align}
Now, we see that only the last exponential factor in  \eqref{e36b} is singular. However,  by virtue of  \eqref{e11} \eqref{e18} and \eqref{e20}, the following  general identity   holds for  $u<u_{in}$:
\be
\label{e37c}
S_{in}(u)+P_{in}^{-1}(u)(\dot P_{in}^{-1}(u))^T=S_{in}(u_0)-\dot P_{in}^{-1}(u)\dot P_{out}(u) (W^T)^{-1}+P_{in}^{-1}(u_0)P_{out}(u_0)(W^T)^{-1};
\ee
moreover,  the right hand side of eq.  \eqref{e37c} is continuous at $u_{in}$. Using this fact in \eqref{e36b} one obtains 
\begin{align}
\label{e37d}
\lim_{u\rightarrow u_{in}^-}  \int e^{-i\bf{q}^T\bx }\Phi_{in}^{k_{v},{\bf k}}&(u,v,\bx) d\bx  =\frac{e^{\frac{-im^2u_{in}}{2k_v}+ik_vv +\frac{\pi i}{2}}}{\sqrt{-4k_v^3\pi |\det( \dot P_{in}(u_{in}))|}}e^{\frac{i}{k_v}{\bf k}^T\dot P_{in}^{-1}(u_{in}){\bf q}} \nonumber\\
\cdot &e^{\frac{-i}{2k_v}{\bf k}^T\left(-\dot P_{in}^{-1}(u_{in})\dot P_{out}(u_{in}) (W^T)^{-1}+S_{in}(u_0)+P_{in}^{-1}(u_0)P_{out}(u_0)(W^T)^{-1} \right){\bf k}}.
\end{align}
Finally, performing the inverse Fourier transform of  the right hand side of eq. \eqref{e37d} and taking into account \eqref{e17b}  we obtain precisely  the formula \eqref{e36}.
\par 
In a  similar way one can consider  the rank one case. However, there appear some technical difficulties (it turns out that eq.  \eqref{e17b}   is valid only in one direction);  thus we present the relevant changes. Let ${\bf v}$ be a null-eigenvector, ${\bf v}^ TP_{in}(u_{in})=0$,  such that ${\bf v}^2=1$.  It is then  easy to check that ${\bf v}=R {\bf e}_2$  where  $R$ is the  orthogonal matrix defined by  eq. \eqref{e36a} and  ${\bf e}_2$ is the second vector of the canonical basis.  Multiplying eq.  \eqref{e37c} by ${\bf v}^TP_{in}(u)$   one obtains
\be
\lim_{u\rightarrow u_{in}^-}{\bf v}^TP_{in}(u)S_{in}(u)=-{\bf v}^T(\dot P_{in}^{-1}(u_{in}))^T.
\ee
On the other hand,  multiplying \eqref{e20} by ${\bf v}^TP_{in}(u)$  one gets
\be
\lim_{u\rightarrow u_{in}^-}{\bf v}^TP_{in}(u)S_{in}(u)W^T=-{\bf v}^TP_{out}(u_{in}).
\ee
Thus, for the rank one case  we have the following counterpart of  the identity  \eqref{e17b}
\be
\label{e37e}
W^{-1}P_{out}^T(u_{in})R{\bf e}_2=\dot P_{in}^{-1}(u_{in})R{\bf e}_2.
\ee
By virtue of  \eqref{e37e} and \eqref{e36a} we see  that the matrix $R^TP_{in}(u_{in})\dot P_{in}^{-1}(u_{in})R$ has  all entries equal to zero      except the first one;  let us  denote it by  $\Theta\in\mR\backslash\{0\} $.  Using  this   and  eq.  \eqref{e22} one obtains     $\lim_{u\rightarrow u_{in}^-} \sign(R^T\dot P_{in}(u)P_{in}^{-1}(u)R)=-1+\sign(\Theta)$. Now, we are  ready  to compute   the Fourier transform of the state $\Phi_{in}^{k_{v},{\bf k}}$  (expressed in terms of ${\bf y}=R{\bf x}$)  and then  perform  the left-sided limit; the result is as follows  
\begin{align}
\label{e37f}
\lim_{u\rightarrow u_{in}^-}  \int e^{-i\bf{q}^T{\bf y} }\Phi_{in}^{k_{v},{\bf k}}&(u,v,{\bf y}) d{\bf y}=\frac{e^{\frac{-im^2u_{in}}{2k_v}+ik_vv +\frac{i\pi}{4}(1+\sign(\Theta k_v))}}{\sqrt{-4k_v^3\pi |\det( \dot P_{in}(u_{in}))|}}e^{\frac{i}{k_v}{\bf k}^T\dot P_{in}^{-1}(u_{in})R{\bf q}}e^{-\frac{i}{2k_v}\Theta (q^1)^2} \nonumber\\
&\cdot e^{\frac{-i}{2k_v}{\bf k}^T\left(-\dot P_{in}^{-1}(u_{in})\dot P_{out}(u_{in}) (W^T)^{-1}+S_{in}(u_0)+P_{in}^{-1}(u_0)P_{out}(u_0)(W^T)^{-1} \right){\bf k}}.
\end{align}
Finally,  making  the inverse Fourier transformation  of  the right hand side of eq. \eqref{e37f} and using  \eqref{e37e} one recovers the   one-dimensional Dirac delta function  obtained above for $\tilde \Phi^{k_v,{\bf k}}$ at  the singular  point  $u_{in}$ (note that  $\sign(\Theta k_v)$ drops out and  there  remains $e^\frac{i\pi}{4} $  factor which coincides  with $e^{\frac{i\pi}{4}(\sign(M(u_0))-\sign(M(u_{in})))}$, cf.  eq. \eqref{e34}).
\par
Finally,   let us compare the above results with the classical picture where the trajectories are   continuous functions on the whole real line. 
To this end, first,   let us recall    that  the   transversal part  of the  geodesic (Lorentz force) equations in the curved spacetime  \eqref{e7} with the  electromagnetic background \eqref{e8} is of the form 
$\overset{..}{\bx}=(K+\frac{2e}{p_v}A)\bx$.  Next, let us note that   there is a  direct relation between  the solutions to the   transversal part of the geodesic equations (the classical description)  and the  solutions to the matrix equation \eqref{e10} (which, in turn, form a basis for the  quantum description). In view of this relation    $\bx(u)=-P_{out}(u)(W^T)^{-1}\frac{{\bf k}}{k_v}+P_{in}(u)\bx_0$ is the   solution of the transversal part of the geodesic (Lorentz force)  equations for the discussed backgrounds   for which  the initial  transversal momenta (i.e. at minus infinity) equal ${\bf k}$ (to see this   eqs.  \eqref{e20a} can be  useful). In consequence,  when $P_{in}(u_{in})=0$   all trajectories  with fixed ${\bf k},k_v$  (but  with various initial parameters $\bx_0$) focus at  the  point $P_{out}(u_{in})(W^T)^{-1}\frac{{\bf k}}{k_v}$ which  coincides with the  Dirac delta behavior of the  of $\tilde \Phi_{in}^{k_v,{\bf k}}$ at $u_{in}$, see eq. \eqref{e36}. The same reasoning can be also applied when the rank of $P_{in}(u_{in})$ is one; then  the second component  $y^2$ of  $y(u)=R^T\bx (u)$  is focused  at the  second component of the   point $\frac{1}{k_v}(R^TP_{out}(u_{in})(W^T)^{-1}{\bf k})$  which  directly  corresponds to  the above quantum results. In particular, when the transversal initial momenta are equal to zero then the focusing takes place at   the point $\bx=0$  or on the line  defined by $y^2=0$, respectively.
 \subsection{The second approach }
 \label{s5}
In view of the above  we find that   after  passing   through  the   singular  point   the phase jump  develops  in the  wave functions describing the ``in" states. This observation  can be confirmed  and extended  to  the case of several  singular points using  the evolution operator. In fact, the dynamic of the field $\Phi_{in}$ is directly determined by   the dynamics of the field  $\bar  \Phi_{in}$; in turn, the evolution of the latter one is governed by that of  a    time-dependent linear oscillator for which the  integral kernel of the evolution operator (propagator) was intensively studied in various approaches, see, among others, Refs.  \cite{b24a}-\cite{b24g}. The propagator for such a system is of   the form 
\be
\label{e37}
U(\bx,u;\bx_0,u_0)=\frac{k_v}{2\pi i \sqrt{|\det( Q_1)}|}e^{\frac{ik_v}{2}(\bx ^T\dot Q_1Q_1^{-1}\bx-2\bx_0^TQ_1^{-1}\bx+\bx_0^TQ_1^{-1}Q_2\bx_0)},
\ee
where $Q$'s are the solutions to eq. \eqref{e10} satisfying   the initial conditions 
\be
\label{e37aa}
 Q_1(u_0)=0, \quad  \dot Q_1(u_0)=I, \quad Q_2(u_0)=I, \quad   \dot Q_{2}(u_0)=0.
  \ee
   A few remarks are in order.  According to the general theory  the singular points of  $Q_1$ and $Q_2$  are isolated; as a consequence   $Q_1^{-1}Q_2$ is a  symmetric\footnote{In fact, by virtue of the initial conditions   $(Q_2^T)^{-1}Q_1^{-1}=\dot Q_1 Q_1^{-1}-(Q_2^T)^{-1}\dot Q_{2}^T$, moreover  the right hand side is a symmetric matrix almost everywhere and $Q$'s are continuous functions.}  matrix (if it  exists).   Moreover,  the    above form of the propagator is valid only when  $\det Q_1(u)\neq 0$ for $u>u_0$  (i.e. when  eq. \eqref{e10} is  disconjugate,   see \cite{b25a, b25b} for more  detailed studies concerning (dis)conjugated equations).  Then,  due to the initial conditions $\det(Q_1(u))>0$ for $u>u_0$  and one can skip the modulus.   However,   as in the case of the ordinary harmonic oscillator,  there  can exist  more singular points   (i.e. eq.  \eqref{e10} is conjugated); and   more careful treatment  of  the evolution operator \eqref{e37} is necessary. This   was intensively studied  in the context of path integral methods,  van Vleck's determinant,  Maslov-Morse's theory, Niederer's transformation, see e.g.  \cite{b14b} and \cite{b24f}-\cite{b24g}.   One   explanation of such a situation  is  based on the observation that the  quantum harmonic oscillator  can be locally related to the free motion and  in order  to obtain the global   propagator the local transformation  have to  be appropriately  glued  -- this  involves  the Maslov-Morse index   (see \cite{b28b,b23f}); more precisely,        the propagator given  by eq. \eqref{e37}  has to be corrected by the  factor of the form  $e^{\frac{i\pi}{2}m(u)}$ (for $k_v<0$)  where $m(u)$   is the number of times counted with multiplicity  such that $\det(Q_1(u))$  is singular since $u_0$. 
To make contact with our reasoning it  is worth to notice that      $\det(Q_1(u))$ vanishes always at $u_0$. Thus conjugated points  (caustics)   are necessary for  the phase jump in the propagator to appear;       
 this is in contrast to the  matrix   $P_{in}$ in    $\Phi_{in}$'s  states,  for which  the  different  initial conditions are imposed,  see eqs. \eqref{e16}.   
\par Let us  now consider the  ``in" states. First,   consider the  disconjugate    case.  Namely, we take  $\bar  \Phi_{in}^{k_v,{\bf k}} (\bx_0,u_0)$  where  the point $u_0$   precedes a singular point,  $u_0<u_{in}$ (note that in  this approach   we do not  need ``out" field  and no  relations  between $u_{out}$ and $u_{in}$ are  required).  
 The  evolution of the state is given by the standard formula $\bar  \Phi_{in}^{k_v,{\bf k}}(\bx,u)=\int U(\bx,u;\bx_0,u_0) \bar  \Phi_{in}^{k_v,{\bf k}}(\bx_0,u_0) d \bx_0$. 
 Now,  using 
 $P_{in}=Q_1\dot P_{in}(u_0)+Q_2P_{in}(u_0)$,  $Q_2^T\dot Q_1-\dot Q_2^TQ_1=I$, $Q_1^T\dot P_{in}-\dot Q_1 ^TP_{in}=-P_{in}(u_0)$,  as well as \eqref{e5},  \eqref{e12}  and \eqref{e25}  we obtain,  after some straightforward but tedious computations, the following result valid   for    $u\geq u_0$ and $u\neq u_{in}$ 
\begin{align}
\label{e37a}
\tilde \Phi_{in}^{k_v,{\bf k}}(u,v,\bx )&=e^{\frac{i\pi}{2}(\frac{\sign(N(u))}{2}+1)}\frac{e^{-\frac{im^2u}{2k_v}+ik_vv+\frac{ik_v}{2}\bx^T\dot P_{in}(u)P_{in}^{-1}(u)\bx+i{\bf k}^TP_{in}^{-1}(u)\bx}}{\sqrt{-2k_v(2\pi)^3}\sqrt{|det(P_{in}(u))|}}\nonumber \\
& e^{-\frac{i}{2k_v}{\bf k}^T(P_{in}^{-1}(u)Q_1(u)(P_{in}^{-1}(u_0))^T+S_{in}(u_0)){\bf k}},
\end{align}
where $N$ is a symmetric  matrix defined as follows  
\be
\label{e39}
N(u)=\frac{k_v}{2}Q_1^{-1}(u)P_{in}(u)P_{in}^{-1}(u_0), \quad  u>u_0.
\ee
Since for $u_0\leq u<u_{in}$ we have 
\be
 \label{e37b}
 S_{in}(u)=P_{in}^{-1}(u)Q_1(u)(P_{in}^{-1}(u_0))^T+S_{in}(u_0),
\ee
   we  can identify   $\tilde S_{in}(u)=S_{in}(u)$ also for $u>u_{in}$; there is no additional constant matrix.  In  consequence, for $u\geq u_0$ and $u\neq u_{in}$ we arrive at the state 
\be
\label{e38}
\tilde   \Phi_{in}^{k_v,{\bf k}}(u,v,\bx)=e^{\frac{i\pi}{2}(\frac{\sign(N(u))}{2}+1)} \Phi_{in}^{k_v,{\bf k}}(u,v,\bx),
\ee
 where  $\Phi_{in}^{k_v,{\bf k}}$,  with    $S_{in}(u) $  given by  the   right hand side of eq.  \eqref{e37b},   makes sense also for  $u> u_{in}$ (this  agrees with our previous conclusions since  it can be shown that  for $u\neq u_{in}$ the  right hand side of eq.   \eqref{e37b} is equivalent to the  right hand side of  eq. \eqref{e34a}).   
The matrix  $N(u)$ apart from being symmetric, obeys     $\sign(N(u))=\sign(N^{-1}(u))$ for $u\neq u_{in}$. Moreover,   in the sufficiently small  neighborhood of $u_0$ the following  expansion holds   $N^{-1}(u)\simeq \frac{2}{k_v}(u-u_0)I$; thus $\sign(N(u))=-2$ for $u\in(u_0,u_{in})$  and,  consequently,   $\tilde   \Phi_{in}^{k_v, {\bf k}}$  coincides with $ \Phi_{in}^{k_v,{\bf k}}$ in this interval.  If  the rank of $P_{in}(u_{in})$ is one then, as we stated above (see eq. \eqref{e22})  $\det(P_{in}(u))$ and, consequently, $\det(N(u))$ change the sign;  thus $\sign(N(u))=0$ for $u>u_{in}$. If the rank of $P_{in}(u_{in})$ is zero then expanding $N(u)$ around $u_{in}$ we find $N(u)\simeq \frac{-k_v}{2}(u-u_{in})(Q_1^TQ_1)^{-1}(u_{in})$ and  $\sign(N(u))=2 $ for $u>u_{in}$.  In consequence, eq.  \eqref{e38} reproduces  our previous result even though  we do not use ``out" states.  
Finally, the Dirac delta behavior of $\tilde   \Phi_{in}^{k_v,{\bf k}}$ at  $u_{in}$ can be also confirmed in this approach by similar reasoning.
\par
Let us now  consider  the conjugated case.  Then it turns out that  we can successively  repeat  the above procedure   to obtain the suitable modification of phase factor  of $\Phi_{in}$.  In fact, let $u_{in}$  be one of the points   where $\det(P_{in})$ vanishes.   Then there exists\footnote{The existence of such a point can be seen  from the observation that $\det(Q^{in}_1) \neq 0$  in a neighborhood of $u_{in}$, where $Q_1^{in}$ is defined  at $u_{in}$ and next using  iii' and iv' of the Theorem 6.3 Sec. V in  \cite{b25a}  } a  point $u_0$,   $u_0<u_{in}$,  such that $Q_1(u_0)=0$  $\dot Q_1(u_0)=I$ and   $\det(P_{in}(u))$ and $\det(Q_{1}(u)) $ do not vanish in an open interval containing  $[u_0,u_{in}]$.  Repeating the above procedure one obtains  the suitable  correction,  ($i$ or $-1$ depending on the  rank of $P_{in}(u_{in})$)  after passing  through  the point $u_{in}$. In consequence, for $k_v<0$,   the change of the phase at each  point $u_{in}$,   such  that $\det(P_{in}(u_{in}))=0, $ is $e^{\frac{i\pi}{2}\textrm{corank}( P(u_{in}))}$ (under our  global assumption that  $ \det(\dot P_{in}(u_{in}))\neq 0 $). Similarly, for the antiparticle state,  $k_v>0$, we have  the correction $e^{-\frac{i\pi}{2}\textrm{corank}( P(u_{in}))}$.
\subsection{Baldwin-Jeffery-Rosen coordinates}
\label{s6}
In this section we  discuss the above results in the context of the Baldwin-Jeffery-Rosen  (BJR) coordinates which are more intuitive for  various  aspects of  the  interaction of the particle with plane  gravitational waves (e.g.  linearized Einstein's equations, the analysis of isometries and integrability of the geodesic equations); in view of this we  skip   electromagnetic field in this section.  Let us recall that  
  the BJR coordinates  $(u, \hat {\bf x },\hat v)$   are related to B coordinates by   the formulae  
\be
\label{e40}
{\bf x}=P\hat {\bf  x}, \quad
  v=\hat v-\frac 1 4\hat {\bf x}^T \dot R \hat {\bf x}, \qquad   R=P^TP,
\ee
 where $P$ is a solution to eq. \eqref{e10}    (satisfying \eqref{e11})  with $A=0$  (then  $P$ does not depend on $k_v$ and \eqref{e40}  define a coordinate  transformation  adopted   to the gravitational field only). 
In the BJR coordinates the plane wave metric takes the form 
 \be
 \label{e41}
  \hat g=2dud\hat v+ d \hat {\bf x}^T Rd \hat {\bf x}, \qquad 
 \ee
 In the weak field approach, in transverse traceless gauge,   the  metric \eqref{e41}  (instead  of \eqref{e3})  is very often a starting point for considerations.  However,  according to the weak energy condition (in particular,  for vacuum solutions) $R$ is singular  at least at one point and thus the metric  \eqref{e41}  is  only a local one;  in order to cover the whole  initial manifold  one  needs at least two  BJR charts (metrics). Moreover, the choice of $R$ (and consequently the localization of singular points) is not fixed a priori; the form of the  metric is invariant under the   transformation  related to the choice of $P$. 
One of the   natural choices seem to be one  based on  $P_{in}$  and $P_{out}$; then we obtain the  ``in"  (``out" respectively)  BJR coordinates   and  metrics. In this approach  the BJR  coordinates tend  asymptotically, at  minus (plus) infinity,  to the Minkowski ones; in other words, the metrics  $\hat g_{in}$  ($\hat g_{out}$) are chosen in such a way that asymptotically they become the Minkowski ones. Moreover,  the asymptotically  free ``in" state $\hat \Phi_{in}^{k_v,{\bf  k}}$ is of the form 
 \be
 \label{e42}
  \hat\Phi_{in}^{{\bf k},k_v}(u,\hat \bx_{in},\hat v_{in})=\frac{1}{\sqrt{-2k_v(2\pi)^3\det(R_{in})}}e^{i\hat v_{in}k_v-\frac{im^2u}{2k_v}} e^{i{\bf k}^T\hat \bx_{in}-\frac{i}{2k_v}{\bf k}^TS_{in}{\bf k}},
 \ee
 where $u<u_{in}$ ($S_{in}$ is the antiderivative of $R_{in}^{-1}$). The inner product in the BJR coordinates can be obtained from \eqref{e9}  (see \cite{b4a}).  
 \par
 First, let us consider the case  when   two maps ``in" and ``out" are  sufficient to cover  the initial B manifold (singular points $u_{in}$ and  $u_{out}$ satisfy $u_{out}<u_{in}$).  Above we have  seen that there is a  phase jump of  the state $\Phi_{in}$  beyond    the singular point $u_{in}$.  This change of phase is reflected in the ``out" region  where  the state  $\Phi_{in}$   is expressed in terms of  ``out" coordinates (on the common domain ($u_{in},u_{out}$) it coincides with $\Phi_{in}$ expressed in terms  of the ``in" BJR coordinates). The situation changes when  both ``in" and ``out" maps are not sufficient to cover the B manifold (for example, when $u_{in}=u_{out}$; see also an example below). Then the change of the  phase appears exactly between both the ``in" and ``out"   BJR maps. Of course, one can choose another ``out" map, e.g. demanding  that its singular point precedes $u_{in}$; however, then the metric   does not have   to be asymptotically a  Minkowski one  (only flat for vacuum solutions).  In the case of more singular points we need more local BJR charts to cover the B manifold and a similar analysis can be performed (see also examples below). 
\section{Explicit examples}
\label{s7}
In this section we  discuss some examples of gravitational and electromagnetic backgrounds,  defined by   \eqref{e7}  and \eqref{e8}, for which   explicit solutions of the K-G   (and  Dirac, see the next section) equation can be found and consequently some  aspects of the interaction  of  quantum fields   with gravitational and electromagnetic waves  can be   analysed more explicitly.  In the literature, especially in the context of gravity,  the case of (linearly polarized) sandwich profiles  was extensively  studied. Here  we shall  analyse some  examples of  continuous   pulses  with non-compact support (including circularly polarized ones).   In view of the   general discussion given in previous sections  the  crucial role plays the   solutions to eq. \eqref{e6}  describing  a   $u$-dependent, linear oscillator. According to the general theory (see e.g. \cite{b29a,b29b})  this  problem   can be reduced to a  classical $u$-dependent linear oscillator.    It turns out that for the latter one there are two  families of continuous profiles    such that the   solutions can be explicitly found and, what is more,  these solutions can be built  from  elementary functions   \cite{b23}.  In consequence,  the quantum dynamics  in the backgrounds defined by these families  can be  studied in more detail.  
\par  Namely, the first family   $g^{(1)}$ of the linearly polarized  plane gravitational waves    is defined by  the profile 
\be
\label{e43}
K^{(1)}(u)=\frac{a}{(u^2+\epsilon^2)^2} G^{(1)}(u), \qquad  G^{(1)}(u)=
\left(
\begin{array}{cc}
1&0\\
0&- 1
\end{array}
\right),
\ee
where $ \epsilon>0$ and $a$ is an arbitrary number (excluding the trivial Minkowski case and redefining, if necessary, $x^1$ and $  x^2$  one can achieve   $a>0$). Moreover, let us note that  taking $a\sim \epsilon^3$  one obtains the model of  an impulsive gravitational wave with  the Dirac delta profile (as $\epsilon$ tends to zero).   
\par 
The second family $g^{(2)}$  provides an example of   the circularly polarized plane  gravitational waves. It is defined by the following profiles 
\be
\label{e44} 
K^{(2)}(u)=\frac{a}{(u^2+\epsilon^2)^2}G^{(2)}(u), \qquad G^{(2)}(u)=
\left(
\begin{array}{cc}
\cos(\phi(u)) &\sin(\phi(u))\\
\sin(\phi(u)) & -\cos(\phi(u))
\end{array}
\right),
\ee
where 
\be
\label{e45}
\phi(u)=\frac{2\gamma}{\epsilon}\tan^{-1}(u/ \epsilon),
\ee
and $\epsilon, \gamma>0$  and  $a$ can be chosen as above. For $\gamma=0$ eq. (\ref{e44})  reduces to the previous case; however,  for physical and mathematical reasons we shall  consider both families  separately. In fact,  the   profiles of the first family  are  even  matrix functions;  moreover, the  quantum dynamics of this family can be  also  explicitly solved   when we change  the sign of the  first entry in $G^{(1)}$; this yields a null-fluid metric.   Such a situation does not hold, in general,  for   the second case.  One can  also consider, in the Minkowski spacetime,  the  suitable electromagnetic potentials $A^{(1)}$  ($A^{(2)}$, respectively)  given by the formulae  \eqref{e44}  and \eqref{e45}. Then  the corresponding electromagnetic fields  satisfy the vacuum Maxwell equations;   even  more, one can combine both   gravitational and electromagnetic backgrounds,  i.e.    $K^{(1)}$  with  $A^{(1)}$ defined by  another   parameter  $a$,  see  eq. \eqref{e10}   (and similarly for the second family). However, to simplify  our notations, we restrict ourselves to  the gravitational case (after  a redefinition of the constant $a$  the  suitable  electromagnetic background  can be included into considerations, cf. eq. \eqref{e10}). Moreover,   from eqs. \eqref{e43} and \eqref{e44} we see  that the profiles fit into  our previous  analysis since $u^2||K^{(i)}(u)||$ belong to the $L^1(\mR)$ space
  and satisfy  eq. \eqref{e15}.     Finally, it is worth to notice that these two families of the fields are distinguished due to the conformal symmetry. Namely, they  exhibit   the 7-dimensional conformal  symmetry, maximal among all (non-flat)  vacuum solutions to the Einstein equations (see \cite{b23b,b23c}); in turn   the electromagnetic fields are invariant under the action of  a special conformal generator \cite{b23dd,b23d}. 
\par 
As we noted above  the classical motion in gravitational (electromagnetic) fields   \eqref{e43} and \eqref{e44} is explicitly  solvable.  Namely, for  $g^{(1)}$      the  solution of  the geodesic equations,   in the  transverse directions\footnote{These directions are the most important since they can be  used to define $P$'s, and to  analyse deviation equations; the explicit form of the $v$ coordinate can be also found,   see \cite{b23}}, for  $a<\epsilon^2$,  is    given by 
\be
\label{e46}
 x^i(u)=C_1^i \sqrt{u^2+\epsilon^2}\sin(\sqrt{\Lambda_i}\tan^{-1}( {u}/{\epsilon})+C_2^i),
\ee
where 
\be
\label{e47}
\Lambda _i={1+(-1)^i\frac{a}{\epsilon^2}} \qquad  i=1,2;
\ee
(for $a>\epsilon^2$  the solution $x^2(u)$ does not change  but    $x^1(u)$ is obtained  by  replacing the   trigonometric functions by their  hyperbolic counterparts).
In consequence, the matrix $P_{in}$  can be explicitly found   and  the  suitable  quantum states $\Phi_{in}$'s  constructed.   
In fact,  taking $C_2^i=\frac{\pi}{2}\sqrt{\Lambda_i}$,  $C_1^1=\frac{1}{\epsilon\sqrt{\Lambda_1}}$ and    $C_1^2=0$  in eq. \eqref{e46}  we  obtain   the  solution $\bx_1(u)$, while taking   $C_1^1=0$  and  $ C_1^2=\frac{1}{\epsilon\sqrt{\Lambda_2}}$,   we obtain another   solution $\bx_2(u)$.    Then,   $P_{in}$ is given by the formula $P_{in} (u)=(\bx_1(u),\bx_2(u))$.  The solution $P_{out}$ can be obtained analogously; alternatively,  by virtue of the fact that  \eqref{e43} is an even  matrix function,  one can put $P_{out}(u)=P_{in}(-u)$.   
 For the second family, the solutions  are more complicated  but  they  can also   be written in terms of elementary functions, see \cite{b23}. 
 \par
Of course, we do not have to refer to classical  solutions to find some solutions of  eq. \eqref{e10}  in such  backgrounds.  The general solution can be found  directly on the quantum level by reducing the initial problem to the well known one.  This alternative approach can be summarized as follows.  For the first family we use the Niederer transformation (see e.g. \cite{b23f,b23e})  to reduce eq. \eqref{e10}  with $K^ {(1)}$ (optionally  with  $A^{(1)}$) to the one describing two separable quantum harmonic oscillators.  More precisely, if  $\bar \Phi$  is a solution of \eqref{e6}   for the first family,   then it is easy to check that 
\be
\label{e48}
\Psi(\tilde  \bx, \tilde u,)=\frac{ \epsilon}{\cos(\tilde u)} e^{\frac{-ik_v\epsilon}{2}\tan(u)\tilde {\bx}^2}\bar \Phi(\frac{\epsilon\tilde  \bx}{\cos(\tilde u)},\epsilon \tan(\tilde u))
\ee
is a solution to   the Schr\"odinger equation  with  potential being the sum of  two independent  harmonic oscillators with frequencies $\Lambda_i$, $i=1,2$, respectively; this observation   leads directly to the solutions of  initial eq. \eqref{e1}. Note that, in general, this transformation is a local one;  its extension is related to the analysis of  the singular points    presented above. 
\par 
For the second family \eqref{e44}  the transformation  \eqref{e48}  leads,     unfortunately,  to  a  $\tilde u$-dependent linear oscillator (it contains trigonometric functions of $\tilde u$).   However,  it is easy to check that  the   global transformation: 
\be
\label{e49}
\Upsilon({\bf y},\tilde u)= \Psi(\textrm{Rot}(\tilde u){\bf y},\tilde u) ,
\ee
where 
\be
\label{e50} 
\textrm{Rot}(\tilde u)=
\left(
\begin{array}{cc}
\cos(\omega \tilde u)&-\sin(\omega \tilde u),\\
\sin(\omega \tilde u)&\cos(\omega \tilde u),\end{array}
\right) , \qquad \omega=\frac{\gamma}{\epsilon},
\ee
define   the state $\Upsilon$ satisfying    the Schr\"odinger equation with  the Hamiltonian $H$ of the form
\be
\label{e51}
H=\frac{{\bf p_y}^2}{2k_v\epsilon}+\frac{k_v\epsilon}{2} (1-\frac{a}{\epsilon^2}){(y^1)^2}+\frac{k_v\epsilon}{2} (1+\frac{a}{\epsilon^2}){(y^2)^2}-\omega(p_{y_2}y^1-p_{y_1}y^2).
\ee
The Schr\"odinger equation  (i.e.  spectrum, eigenfunctions) of  the  Hamiltonian of this   type  has been considered in many  papers   and  in various approaches, see e.g. \cite{b30a,b30,b30b}. Here, we use the reasoning  presented in \cite{b30} and define a unitary operator $V$: 
\be
\label{e52}
V=e^{i\alpha y_1y_2}e^{i\gamma p_{y_1}p_{y_2}} ,
\ee
where $\alpha$ and $\gamma$ are constants adjusted in such a way  that   $V^+HV$ is separable; these constants   can be easily expressed in terms of  the know parameters ($k_v,a,\epsilon $) by means of eqs.  (6) and (7)  from  \cite{b30}.  Then for  the parameters $|\epsilon^2-\gamma^2|>a$ the transformed Hamiltonian   $V^+HV$ is  equal to   the sum of two independent Hamiltonians  describing  harmonic oscillators\footnote{It seems that  there is a printing error    in  the formula for  $H_3$  in \cite{b30}; namely   instead of $(c-2\alpha a)W$   there should be $(c+2\alpha a)W$.};  they frequencies are as follows 
\be
\label{e53}
\begin{split}
\Omega_1^2={1+\omega^2+\sqrt{4\omega^2+\Omega^2}},\\
\Omega_2^2={1+\omega^2-\sqrt{4\omega^2+\Omega^2}};
\end{split}
\ee
where 
 \be
 \label{e54}
  \Omega=\frac{a}{\epsilon^2}.
 \ee
These results  exactly agree with    the classical ones  (see \cite{b23}).   Now, the eigenfunctions for $H$ can be built by means of  the operator  $V$  and the  well known Hermite functions. However,  since the operator $V$ contains momenta  the computations  are involved.  The final result, see  eq. (13) in  Ref. \cite{b30}, is that  the eigenfunctions for $H$,  given by  eq. \eqref{e51},  contain  finite sums of products of two Hermite polynomials with different complex arguments.  Now, performing  the  inverse transformations to \eqref{e48}  and \eqref{e49} we can construct explicit solutions of the K-G  eq. \eqref{e1}   in the gravitational field $g^{(2)}$  (and, optionally, including the electromagnetic potential $A^{(2)}$, also in the Minkowski space). Moreover,  we  have shown above that the quantum cross section coincides with its classical counterpart; the latter one was explicitly found for both families of backgrounds \cite{b23}  thus we immediately obtain its   quantum counterpart.   
\par 
Finally, let  us  go back to singularities.   For the profile $K^{(1)}$  the zeros of $\det(P_{in})$ are determined by  the second diagonal component  of  $P_{in}$. For $\Lambda_2<2$  there  is one singular  point $u_{in}>0$   and thus two maps are sufficient to cover the whole manifold  (since  $K^{(1)}$  is an  even function,   $P_{out}(u)=P_{in}(-u)$ and $u_{out}=-u_{in}<0$). For $\Lambda_2=2 $  (equivalently, $a=3\epsilon^2$) there is   again  one  singular point $u_{in}=0$ (the second  component of $P_{in}$   is given by  $-u/\sqrt{u^2+\epsilon^2}$), but this time  $u_{out}=u_{in}$; thus,   the maps related to $P_{in}$ and $P_{out}$ are not sufficient to cover the B manifold. Of course, one can  take instead of $P_{out}$  another $P$  for  which the  determinant  vanishes at some negative point. Then  together  with $P_{in}$ they cover the B manifold; however,  at plus infinity  the BJR  metric do not coincide with the Minkowski one. For $\Lambda_2>2$ the situation gets complicated and  the number zeros growths; in consequence, the number of phase corrections and BJR maps growths too. Since  for all these points the rank of $P_{in}$ is one  the  suitable phase correction is always  $e^{i\frac{\pi}{2}}$.  Similar situation holds for the second family; then, however,    the singular points are given  more implicitly,  see \cite{b23}. 
\section{Fermionic fields}
\label{s8}
In this section we  analyse  all the  above issues for a massive fermionic   fields.  
To this end  let us recall that the  the coupling of    spin one-half fields    to the  metric $g$ and external electromagnetic potential $A_\mu$   is described  by the   spinor field $\Psi$ satisfying the  Dirac equation 
\be
\label{e61}
({\gamma}^{\mu}(\partial_{\mu}-\Gamma_{\mu}-ie A_{\mu}I)+mI)\Psi=0,
\ee
where $ \gamma^{\mu}\equiv g^{\mu\nu} \gamma_{\nu}$, $ \gamma_{\mu} \equiv {e^{\bar  \nu}}_{\mu} \gamma_{\bar \nu}$, 
\be 
\label{e62}
\Gamma_\mu= -\frac 1 4 \gamma_{\bar  \mu}\gamma_{\bar \nu}{e^{\bar \mu}}_{\nu}g^{\nu\alpha}(\partial_\mu {e^{\bar\nu}}_{\alpha}-\Gamma^\beta_{\mu\alpha}{e^{\bar \nu}}_{\beta}),
\ee 
and   the orthonormal tetrad fields are  defined by  the condition $g_{\mu\nu}=\eta_{\bar \mu\bar \nu} {e^{\bar \mu}}_{\mu}{e^{\bar \nu}}_\nu$ ($\gamma^{\bar \mu}$  are the ordinary gamma matrices in the Minkowski spacetime $x^{\bar \mu}$ with $\eta_{\bar\mu\bar\nu}=(-,+,+,+)$ i.e.,  $\gamma^{\bar 0}=-\gamma_{\bar 0}=-i\beta$ and $\vec\gamma=-i\beta\vec\alpha$). 
\par 
For the  pp-wave  metric $g$  defined  by \eqref{e3} the  tedrad ${e^{\bar \mu}}_{\mu}$  can be easily  found together with the  matrix functions  $\gamma_{\mu}$. Then  one  gets the   matrices $\Gamma_\mu$:
\be
\label{e63}
\Gamma_u=\frac{1}{8\sqrt 2}\partial _ k\mK [\gamma^{ k},\gamma_{\bar 0}+\gamma_{\bar 3}],\quad \Gamma_v=0, \quad \Gamma_k=0.
\ee
Now, introducing  new 2-components spinorial fields   $\Psi_u$ and $\Psi_v$ by an invertible (unitary up to a constant factor)  transformation 
\begin{align}
\label{e64}
\left(
\begin{array}{cc}
\Psi_u\\
\Psi_v
\end{array}
\right)
\equiv \left(
\begin{array}{cc}
\sigma_3&-I\\\
I&\sigma_3
\end{array}
\right)\Psi,
\end{align}
one finds that  the Dirac equation \eqref{e61},  in the pp-wave  and electromagnetic   backgrounds \eqref{e4},  is equivalent to the following set of equations
\begin{align}
\label{e65a}
i\sqrt 2 \partial_u\Psi_u&=-e\sqrt 2 \mA\Psi_u+\frac{i\mK}{\sqrt 2}\partial_v\Psi_u-(i\sigma_k\partial_k+m\sigma_3)\Psi_v\\
\label{e65b}
i\sqrt 2 \partial_v\Psi_v&=(i\sigma_k\partial_k+m\sigma_3)\Psi_u.
\end{align}
Since the coefficients in the above set of equations do not depend on $v$  we look for solutions of the form (cf. \eqref{e5})
\be
\label{e66}
\Psi_u=e^{{i vk_v  -i\frac{m^2u}{2k_v}}}\bar \Psi_u,  \qquad  \Psi_v=e^{{i vk_v  -i\frac{m^2u}{2k_v}}}\bar \Psi_v.
\ee
Substituting this ansatz into  eqs. \eqref{e65a} and \eqref{e65b} one concludes that the  equation for the spinor field $\bar \Psi_u$ decouples and   its both components satisfy  the same    Schr\"odinger equation (cf. \eqref{e6}), namely  
\be
\label{e67}
i\partial_u\bar\Psi_u=-\frac{1}{2 k_v}\triangle_\perp \bar\Psi_u-\frac{k_v}{2}(\frac{2e}{k_v}\mA+\mK)\bar \Psi_u;
\ee
moreover, the 2-components field  $\bar \Psi_v$  (and thus  $\Psi_v$) is directly determined by $\bar \Psi_u$ 
\be
\label{e68}
\bar \Psi_v=-\frac{1}{\sqrt 2 k_v}(i\sigma_l\partial_l+m\sigma_3)\bar\Psi_u.
\ee
\par Now, as in the   spinless case,  we restrict  ourselves to the backgrounds defined by \eqref{e7} and \eqref{e8}; then   the solutions of   the Dirac equation  \eqref{e61}    can be directly constructed   in terms  of  the bosonic fields \eqref{e12}. Namely,   the component    $\Psi_u^{k_v,{\bf k},{\bf w}}$  (compare eq.  \eqref{e6}  and \eqref{e67})  takes the form 
  \be
  \label{e69}
\Psi_u^{k_v,{\bf k},{\bf w}}=\sqrt{2|k_v|}\Phi^{k_v,{\bf k}}{\bf w} ,
\ee
where ${\bf w}$ is a two-dimensional   vector. The final form of the fermionic state reads 
\be
\label{e70}
\Psi^{k_v,{\bf k},{\bf w}}= \sqrt{\frac{|k_v|}{2}}\Phi^{k_v,{\bf k},{\bf w}}
\left(
\begin{array}{cc}
(1-\frac{m}{\sqrt 2 k_v})\sigma_3{\bf w}+\frac{1}{\sqrt 2 k_v}({{ \partial_k}} f { \sigma_k}){\bf w}\\
-(1+\frac{m}{\sqrt 2 k_v}){\bf w}-\frac{1}{\sqrt 2 k_v}({{ \partial_k}} f { \sigma_k\sigma_3}){\bf w}\\
\end{array}
\right),
\ee
where $f={\bf k^T}P^{-1}{\bf x}+i\frac{k_v}{2}{\bf x}^T\dot P P^{-1}{\bf x}$ (for the linearly polarized gravitational waves    it reduces to  one obtained in Ref.  \cite {b6} in local  BJR coordinates).
\par 
Let us introduce the  inner product  as follows 
\be
\label{e71}
\braket {\Psi_1}{\Psi_2}=\int\Psi_1^\dag \gamma^{\bar 0}\gamma^u\Psi_2dvd{\bf x}.
\ee
It is worth to notice that $\Pi=\frac{1}{\sqrt2}\gamma^{\bar 0}\gamma^u$ is a projector operator and, consequently, the inner product defined by \eqref{e71} depends   on the $u$-spinorial component only  
\be
\label{e72}
\braket {\Psi_1}{\Psi_2}=\int \frac{1}{\sqrt{2}}\Psi_{1u}^\dag \Psi_{2u} dvd{\bf x}.
\ee
In view of  this the  solutions $\Psi^{k_v,{\bf k},k}$   (see \eqref{e69}) defined by the vectors ${\bf w}_k$, $k=1,2$    of the form
\be
\label{e73}
{\bf w}_k=2^{\frac 14}{\bf  e}_k, \quad k=1,2;
\ee
where $(e_k)_{k=1}^2$ is the standard 2-dimensional  canonical basis,  form an orthonormal set of functions with respect to the  inner  product \eqref{e71}
\be
\label{e74}
\braket {\Psi^{k_v,{\bf k},k}}{\Psi^{l_v,{\bf l},l}}=\delta(k_v-l_v)\delta^{(2)}({\bf k}-{\bf l})\delta_{kl}.
\ee
Moreover, the states ${\Psi^{k_v,{\bf k},k}}$ with  $k_v\in\mR\backslash \{0\}$, ${\bf k}\in \mR^2$ and   $k=1,2$, form   a complete set. This can be directly shown by taking into account  the projector operator $\Pi$   and the form of $\Psi$'s (eq. \eqref{e70}).  As we noted above, for  the bosonic case and  the   gravitational  backgrounds  only,  as a complete basis we can take the states $  {\Phi^{k_v,{\bf k}}}$ and  $ {\Phi^{k_v,{\bf k}}}^*$  with negative $k_v$, this implies a similar situation for the state $\Psi$'s; however, now  for the states with positive $k_v$  there are also  some sign changes in   the  spinorial  part  of  \eqref{e70}, namely the replacement $m \rightarrow -m$ should be made.
\par In Sec. \ref{s7}  we noted that   the matrices $P$ (and consequently the bosonic  states $\Phi$) can be explicitly found  for the two  gravitational    families defined by the profiles  \eqref{e43} and \eqref{e44} (optionally with  the suitable electromagnetic backgrounds).      In view of eq. \eqref{e70}  the same   holds also   for the spin one-half particle described by the Dirac equation in these  backgrounds. Finally,   let us  also  note  that when we switch off the   gravitational and electromagnetic fields (and put $P=I$)    the states  $\Psi^{k_v,{\bf k},k} $ become equivalent  to the ones  for the Minkowski spacetime     known from the  standard  textbooks.   
\par
Now, as in the case of  bosonic fields   we   define, in terms   of the   matrix $P_{in}$ ($P_{out}$), the states   $\Psi^{k_v,{\bf k},k}_{in}$ (respectively ``out" states)  which asymptotically tend to the free ones. Then,  using eq. \eqref{e72}  we can  compute the inner product between them  and  the transition  amplitude from an ``in" one-particle state to  an ``out" one-particle state 
\be 
|\braket {\Psi^{k_v,{\bf k},k}_{in}}{\Psi^{l_v,{\bf l},l}_{out}}|=\frac{\delta(k_v-l_v)}{2\pi |l_v|\sqrt{|\det(W)}|} \delta_{kl}.
 \ee
 Summing over all the final spin  states and then taking averaging over the initial polarization  one  obtains the final  quantum cross section which is  of   the same form  as in the bosonic case, i.e.   given  by  eq. \eqref{e30} (extending  in this way the results of   Ref. \cite{b6}  obtained  for fine-tuned and  linearly polarized  plane gravitational sandwiches). In particular,   we immediately obtain the explicit form of the  cross section for the backgrounds defined by \eqref{e43} or \eqref{e44} (see Sec. \ref{s7} and \cite{b23}). 
\par
Moreover,   as for the bosonic field $\Phi^{k_v,{\bf k}}_{in}$, there is a phase change  for the fermionic  fields  after passing  through the singular point.  Namely,  the  $u$-component of the field $\Psi^{k_v,{\bf k},k}_{in}$ is proportional to $\Phi^{k_v,{\bf k}}_{in}$ (cf.  eq. \eqref{e69}); thus  it carries the same correction. In view of eq. \eqref{e68}  this concerns also the $v$-component.  As a result  the  phase of the    full  state  $\Psi^{k_v,{\bf k},k}_{in}$ also changes after passing  through the singular point. 
\par
Finally,   let us analyse  the change of the spin polarization. It is difficult to  define the spin operator  for  a general   curved spacetime; however, in  the Minkowski spacetime in the  rest frame    it can be  easily described by  three  matrices 
\be
\vec \Sigma=\left(
\begin{array}{cc}
\vec \sigma&0\\
0& \vec \sigma\\
\end{array}
\right).
\ee
Thus we can try to analyse the change of the spin  polarization    after  the pulse has passed  (where   the metric is again the Minkowskian one). Such  an approach was applied (in the  Newman-Penrose  framework)   in Ref.  \cite{b7}  to the linearly polarized sandwich waves; here we study  this problem, in  a slightly different way, for the general plane sandwich waves (for example, one  can  take the pulses studied  in   Ref.  \cite{b31}).  In such a case  the Minkowskian  region emerges for a sufficiently large   parameter $u$.  We start with the  rest particle  state, i.e.  ${\bf k}={\bf 0}$,  $k_v=\frac{-m}{\sqrt 2}$, at minus null infinity.   Then, at a neighborhood of  minus infinity,  the  positive energy  spinor wave functions  $\Psi_{in}^{\frac{-m}{\sqrt 2}, {\bf 0}, {\bf w}_a} $,  $a=1,2,3$      defined by the vectors  of the form
\be
{\bf w}_1=
\left(
\begin{array}{c}
1\\
-1
\end{array}
\right),
\quad
{\bf w}_2=
\left(
\begin{array}{c}
1\\
-i
\end{array}
\right),
\quad 
{\bf w}_3=
\left(
\begin{array}{c}
1\\
0
\end{array}
\right),
\ee
  are   eigenfunctions  (with the eigenvalue $1$) of the matrix $\Sigma_a$,  respectively;  i.e. their spins are  polarized in the three  directions. Now let us analyse  the "after" Minkowskian  region where the metric is again  the Minkowskian one.     To this end, in this region,  we act   with an arbitrary (i.e. defined by arbitrary momenta $\vec{p}$)  Lorentz  boost in spinor representaion   on the   field $\Psi_{in}^{\frac{-m}{\sqrt 2}, {\bf 0}, {\bf w}_a} $  and then  check, by direct calculations, that the  field   obtained in this way  is  not  an eigenfunction (with the initial eigenvalue)  of the  matrix $\Sigma_a$.  In consequence there is no rest  frame such that  the spin polarization  is the same as the initial one (the effect of the gravitational wave pulse is that the Dirac particle loses its initial polarization see also \cite{b7}). Thus, there is a change of the spin polarization (a kind of the spin memory effect, cf. \cite{b11})  after passing  the general sandwich  wave.
 \section{Conclusions and outlook}
 \label{s9}
 In this paper we have discussed   some aspects of    massive quantum fields in the   general  plane gravitational wave in the  presence  of non-plane electromagnetic waves; in particular, the  asymptotic conditions, ``in" (``out")  states   and  the  cross sections were  obtained. We observed  that    despite of singularities of  these  states (corresponding, on the classical level, to   the focusing properties)  their    global form  can be established:  at the singular points  the Dirac delta (distributional)   behavior emerges  and there is a phase  jump of the wave function  after passing  through   each  singular point.    
 Moreover, we noted that  this phase correction  can be described  in terms of  relations  between   BJR charts.  We  also discussed   some  examples  of the waves  of infinite range for which the explicit form of the states can be obtained (in particular,  the circularly polarized ones).  All these results concern both the  scalar  as well  as spin one-half  quantum fields; moreover, in the latter case we analysed the change of  the spin polarization after passing  the  general  sandwich wave.
 \par
Turning to possible further developments let  us recall that the  Penrose limit \cite{b10}  of the spacetime yields a plane gravitational wave. In consequence,   the propagation of quantum fields  in a more realistic spacetime  can be approximated  in some local region (near  the null geodesics)  by considering    an appropriate  plane wave spacetime. Such an  approach was  successively applied in  Refs.  \cite {b16}-\cite{b18c} to   the Green  functions  and then  vacuum polarization  in QED.  In this context, it would be also  interesting to describe spacetimes for  which  the explicitly solvable   examples   presented in Sec. \ref{s7} emerge as the   Penrose limit.    The other quite natural  generalization  corresponds to going beyond the plane waves and considering some pp-waves,    gravitational wave beams \cite{b32} and  the linearized theory  \cite{b33}.  It is also worth to notice that the  case of   massless     fields  involves also  a special  analysis   due to the  possibility of  massless  particles creation  by the gravitational field  (in particular,  see the recent  considerations   in \cite{b22a,b22b});   actually,   such a possibility  can exist even for the  massive fields; however, then some additional constrains  have to be added  \cite{b34}.    Moreover,  the  results obtained    may be useful  in the context of   trapping and soft  problems in gravity \cite{b8c,b35a,b35b}.  Finally, following  Refs. \cite{b36a}-\cite{b36d} we hope that they can be also  useful in the study of the light-matter interaction. 
   
\vspace*{0.3cm}
\par
{\bf Acknowledgments}\\
Comments of Cezary  Gonera, Joanna  Gonera and Pawe\l\      Ma\'slanka are   acknowledged.  We are also  grateful to the anonymous referees  for   valuable remarks which certainly  improved  the paper. The research has been supported by the grant 2016/23/B/ST2/00727 of National Science Center, Poland. 

 \end{document}